\documentclass[preprint]{revtex4-1} 

\usepackage{amsfonts} 
\usepackage{graphicx}
\usepackage{amsmath, amssymb}
\usepackage{slashed}
\usepackage{braket}
\usepackage{physics}
\usepackage[bottom]{footmisc}
\graphicspath{  {images/} }
\begin{document}


\title{Relativistic Equations for Fractional-Spin particles}

\author{Satish Ramakrishna}
\email{ramakrishna@physics.rutgers.edu}
\affiliation{Department of Physics \& Astronomy, Rutgers, The State University of New Jersey, 136 Frelinghuysen Road
Piscataway, NJ 08854-8019}


\date{\today}

\begin{abstract}

This paper generalizes the method of deducing Dirac's equation to constructing a family of equations that represent the $N$-th root of the basic Energy-Momentum relation for a free particle \cite{Dattoli1, Dattoli2, Dattoli3, Dattoli4}. Then these equations are recast in a form that allows one to interpret them as the fundamental dynamical equations for particles with fractional spin, which we study in detail for the case of  $N=4$ and $N=3$. We explicitly prove that the equation is invariant to rotations and boosts and indeed represents spin-$\frac{3}{8}$ and spin-$\frac{1}{8}$ particles for $N=4$ and spin-$\frac{1}{6}$ and $0$ for $N=3$.

\end{abstract}
\maketitle 

Dirac's original method of deriving a novel equation for spin-$\frac{1}{2}$ particles was, to put it informally, a square root of the basic Energy-Momentum relationship of a free particle. In the process, the wave-function describing the particle increases in dimensionality. This allows one to describe the behavior of spin-$\frac{1}{2}$ particles. The equation's structure affects the coupling of the particles to magnetic fields etc. and produces dynamical effects that are special to fermions.

This paper was meant to explain how to generalize the method to take higher roots (than the square root). While it was being written, I  became aware that the mathematical idea had been explored, exactly as I had done, earlier \cite{Dattoli1,Dattoli2,Dattoli3,Dattoli4}. However, the analysis presented here allows one to deduce, further, that the spin of the particles involved in $2$-dimensions are indeed fractional, as expected for anyonic excitations. The equations that describe the wave-function turn out to involve fractional derivatives of space and time. We study, in particular, the case $N=4$ and $N=3$ and prove that the equation describes spin-$\frac{3}{8}$ and spin-$\frac{1}{8}$ particles (for $N=4$) and spin-$\frac{1}{6}$ and spin-$0$ (for $N=3$).
 
 The plan of the paper is as follows. We begin with a recap of the application of the method to the original case in $2+1$-dimensions. We write the equation in four equivalent bases and demonstrate covariance to rotations and boosts in these bases. Then we deduce the equation that results from one of the bases.
 
 The same approach is used to deduce different bases for the case $N=4$; we use these bases to deduce covariance under rotations and boosts. Again, we deduce the equation that describes the time-evolution of the wave-function. We generaliize the idea to higher even $N$ and deduce the spins of the particles represented by those equations.
 
 Then we apply a similar technique to deduce results for $N=3$, prove the covariance under rotations and boosts, as well as generalize the idea to higher odd $N$.
 
 Finally, we show, in one example, that the time evolution is unitary.
 
\section{Recap of Dirac's method, $N=2$}

We start with the energy-momentum relation in $2+1$ dimensions
\begin{eqnarray}
m^2= E^2 - p_1^2 - p_2^2
\end{eqnarray}

Dirac's method was to find a linear representation of the above equation (ostensibly a ``square-root'')
\begin{eqnarray}
{\cal I}_2 m =  \sigma_3 \:  E - i \sigma_2 \: p_1 + i \sigma_1 \: p_2  
\end{eqnarray}
and in order to match the energy-momentum relation by squaring Equation(2), we need to impose
\begin{eqnarray}
\{\sigma_i, \sigma_j \} = 2 \delta_{ij} 
\end{eqnarray}
In line with a description we will use ahead in the paper, we refer to these as 2-anticommuting matrices. In the above ${\cal I}_2$ is the $2 \times 2$ identity matrix.

A well-known and convenient solution to the conditions in Equation (3) is  (with $z_2=e^{i \pi}=-1$)
\begin{eqnarray}
\sigma_1 =
\left(\begin{array}{cc} 0 & 1 \\
			   1 & 0 \end{array}\right) \: \: \: 
\sigma_2 =\left(\begin{array}{cc} 0 & -i \\
			   i & 0 \end{array}\right)  \: \: \: 
\sigma_3 \equiv Q_2 = 
\left(\begin{array}{cc} {\bf 1} &  0 \\
			0 & z_2 \end{array}\right)
\end{eqnarray}

In matrix form, the above Equation (2), written as an operator acting on a state, is
\begin{eqnarray}
{\cal A}_2 \left(\begin{array}{c} \psi_1 \\
			\psi_2 \end{array}\right) \equiv
\left(\begin{array}{cc} E-m & - p_- \\
			 p_+ & -(E+m )\end{array}\right) \left(\begin{array}{c} \psi_1 \\
			\psi_2 \end{array}\right) = 0
\end{eqnarray}

We can show that the matrix equations above are invariant to rotations and boosts, with appropriate transformations. 
We write the above equation in two other bases. To be aligned with the discussion ahead, we will label them $1,2,3,4$.
\begin{eqnarray}
Basis \: 1: \: \: \: \: \: m {\cal I}_2 = Q_2 \: E + \left[ (-i \sigma_2) p_1 + i \sigma_1 p_2 \right] \: \: \: \: \: \: \: \: \: \: \: \: \: \: \:\: \: \: \: \:\: \: \: \: \:\: \: \:   \nonumber \\
Basis \: 2: \: \: \: \: \: m {\cal I}_2 = Q_2 \: E + \left[ -  T_{2+} p_- +  T_{2-} p_+ \right] \: \: \: \: \: \: \: \: \: \: \: \: \: \:\: \: \: \: \:\: \: \: \: \:\: \: \:    \nonumber \\
Basis \: 3: \: \: \: \: \: \: \: \: \: \:m{\cal I}_2 = Q_2 \: i p_2 -  \left[  T_{2+} (E+p_1) +  T_{2-} (E-p_1) \right] \nonumber \\
Basis \: 4: \: \: \: \: \: m {\cal I}_2 = - Q_2 \: i p_1 \: - \left[ -T_{2+} (E+p_2) +  T_{2-} (E-p_2) \right]
\end{eqnarray}
These bases can be derived as follows - (2) can be derived from (1) by algebraic manipulation. (3) and (4) are different basis representations of the equations represented by bases (1) and (2).

To demonstrate the invariance to rotations, using exactly the same procedure that will be used in the $N=4$ case, we use the operator ${\cal R}_2$ below and apply it to ``Basis 2'' above. In particular, under rotations, the operator ${\cal R}_2$ renders the equation invariant, i.e., applying ${\cal R}_2^{-1}(...){\cal R}_2$  leaves the form of the equation in ``Basis 2'' unchanged.
\begin{eqnarray}
p_{-} \rightarrow p_{-} e^{-i \theta} \: \: \: \: \: \: \: \: \: \: \: 
\: \:  \: \: \: p_{+} \rightarrow p_{+} e^{i \theta} \: \: \: \: \: \: \: \: \: \: \: \: \: \: \: \: \: 
{\cal R}_2 = \left(\begin{array}{cc} e^{-\frac{i}{2} \theta} \: \: \: & 0  \\
						0 & e^{\frac{i}{2} \theta} \: \\  \end{array}\right)	
\end{eqnarray}

Under boosts, for instance along the $1$-direction, we combine terms as follows to derive the equation labeled ``Basis 3'':
\begin{eqnarray}
\sigma_A (E + p_1) + \sigma_B (E-p_1) + (i \sigma_1) p_2 = m  \: \: \: \: \: \: \: \: \: \: \: \:  \: \: \: \: \: \: \: \: \: \: \: \: \: \: \: \: \: \: \: \: \: \: \: \: \: \: \: \: \: \: \: \:  \nonumber \\
\sigma_A = \frac{1}{2} (\sigma_3-i \sigma_2)= \frac{1}{2} \left(\begin{array}{cc} 1 \: \: \: & -1  \\
						1\: \: & -1 \: \\  \end{array}\right) \: \: \: , \: \: \: \sigma_B = \frac{1}{2}(\sigma_3+i \sigma2) =  \frac{1}{2} \left(\begin{array}{cc}  \: \:  1&  \: \: 1  \\
						-1 & -1 \\  \end{array}\right) \: \: \: \: \: \: \: \: \: \: \: \:  \: \: \: \: \: \: \: \: \: \: \: \: \nonumber \\
\sigma_A^2=\sigma_B^2 = 0 \:  , \:  (i \sigma_1)^2 = -{\cal I}  \:  , \:  \{\sigma_A,\sigma_B\}={\cal I} \:  , \:  \{\sigma_A,i \sigma_1\}=\{\sigma_B, i \sigma_1\} = 0 \: \: \: \: \: \: \: \: \: \: \: \: \: \: \: \: \: \: \: \: \: \: \: 
\end{eqnarray}
The equation is identical to the earlier equation and indeed, squaring both sides yields the usual energy-momentum relation.  Next, we transform to a basis that diagonalizes $i \sigma_1$. The diagonalizing matrix $L_2$ is
\begin{eqnarray}
L_2= \left(\begin{array}{cc} 1 & -1  \\
						1 & 1  \\  \end{array}\right) \: \: \: , \: \: \: L_2^{-1}= \frac{1}{2} \left(\begin{array}{cc} 1 & 1  \\
						-1 & 1 \\  \end{array}\right) 
\end{eqnarray}
and we find an equation with a non-Hermitian operator
\begin{eqnarray}
L_2 \sigma_A L_2^{-1} \:  (E+p_1) + L_2 \sigma_B L_2^{-1} \:  (E-p_1) +  \left(\begin{array}{cc} i \: \: \: & 0  \\
						0 & -i \: \\  \end{array}\right)	 p_2 = m \:  {\cal I} \nonumber \\
\rightarrow  \left(\begin{array}{cc} 0 \: \: \: & 0  \\
						1 & 0 \: \\  \end{array}\right)	 (E-p_1) + \left(\begin{array}{cc} 0 \: \: \: & 1  \\
						0 & 0 \: \\  \end{array}\right)	 (E+p_1) + \left(\begin{array}{cc} i \: \: \: & 0  \\
						0 & -i \: \\  \end{array}\right) p_2 =m \: {\cal I} 
\end{eqnarray}

In this basis, the boost matrix is 
\begin{eqnarray}
(E+p_1) \rightarrow (E+p_1) e^{t} \: \: \: , \: \: \: (E-p_1) \rightarrow (E-p_1) e^{-t} \: \: \: \: \: \: \: 
{\cal B}_2 = \left(\begin{array}{cc} e^{-\frac{1}{2} t} \: \: \: & 0  \\
						0 & e^{\frac{1}{2} t} \: \\  \end{array}\right)	
\end{eqnarray}
and applying the transformation ${\cal B}_2^{-1} (...) {\cal B}_2$, we find the boost leaves the equation invariant.
Then, reverting back to the original basis, the boost matrix would be
\begin{eqnarray}
{\cal B}_2^{orig} = L_2^{-1} {\cal B}_2 = \frac{1}{2} \left(\begin{array}{cc} e^{-\frac{1}{2} t} \: \: \: &  e^{\frac{1}{2} t}  \\
						-e^{-\frac{1}{2} t} & e^{\frac{1}{2} t} \: \\  \end{array}\right)	
\end{eqnarray}
A similar calculation can be performed for the boost operator along the $2$-direction.

To derive an equation for wave-function from this formalism, the recipe is to make the replacements $E \rightarrow i \hbar \frac{\partial }{\partial t}, p_i \rightarrow -i \hbar \frac{\partial }{\partial x^i}$ and get the Dirac equation in 2-dimensions, which is, in matrix form
\begin{eqnarray}
\left(\begin{array}{cc} i \hbar \frac{\partial }{\partial t} \: - \: m & i \hbar (\frac{\partial}{\partial x}- \frac{\partial }{\partial y}) \\
			 i \hbar (\frac{\partial}{\partial x}+ \frac{\partial }{\partial y}) &  i \hbar \frac{\partial }{\partial t} \: + \: m \end{array}\right) \left(\begin{array}{c} \psi_1 \\
			\psi_2 \end{array}\right) = 0
\end{eqnarray}

In the above, we have studied the case of $N=2$, we will shortly generalize this to higher values of $N$. We begin by studying the case $N=4$.

\section{Generalizing to higher dimensional matrices and even-fractional powers of variable, N=4}

Starting with the energy-momentum relation, we write (${\cal I}_4$ is the $4 \times 4$ identity matrix) in the following four different ways
\begin{eqnarray}
Basis \: 1: \: \: \: \: \: m^{\frac{1}{2}} {\cal I}_4 = Q_4 \: E^{\frac{1}{2}} + \sqrt{i} \left[ B p_1^{\frac{1}{2}} + A p_2^{\frac{1}{2}} \right] \: \: \: \: \: \: \: \: \: \: \: \: \: \: \:\: \: \: \: \:\: \: \: \: \:\: \: \:   \nonumber \\
Basis \: 2: \: \: \: \: \: m^{\frac{1}{2}} {\cal I}_4 = W_4 + \left[  T_{4+} p_-^{\frac{1}{2}} +  T_{4-} p_+^{\frac{1}{2}} \right] \: \: \: \: \: \: \: \: \: \: \: \: \: \:\: \: \: \: \:\: \: \: \: \:\: \: \:    \nonumber \\
Basis \: 3: \: \: \: \: \: m^{\frac{1}{2}} {\cal I}_4 = Y_4 + i \left[  T_{4+} (E-p_1)^{\frac{1}{2}} +  T_{4-} (E+p_1)^{\frac{1}{2}} \right] \nonumber \\
Basis \: 4: \: \: \: \: \: m^{\frac{1}{2}} {\cal I}_4 = Z_4 \:+ i \left[  T_{4+} (E-p_2)^{\frac{1}{2}} +  T_{4-} (E+p_2)^{\frac{1}{2}} \right]
\end{eqnarray}
In the above, the matrices $A_4, B_4, Q_4$ are of a type we refer to as $4$-anticommuting, i.e., the following matrix equations are obeyed \cite{Dattoli1}
\begin{eqnarray}
A_4.A_4.A_4.A_4={\cal I} \: \: \: , \: \: \: B_4.B_4.B_4.B_4={\cal I} \: \: \: , \: \: \: Q_4.Q_4.Q_4.Q_4={\cal I}
\end{eqnarray}
and
\begin{eqnarray}
A_4.A_4.A_4.B_4+A_4.A_4.B_4.A_4+A_4.B_4.A_4.A_4+B_4.A_4.A_4.A_4=0 \: \: \: \: \: \: \: \: \: \: \: \: \: \: \: \: \: \: \: \: \: \: \: \: \: \: \: \: \: \: \nonumber \\
A_4.A_4.A_4.Q_4+A_4.A_4.Q_4.A_4+A_4.Q_4.A_4.A_4+Q_4.A_4.A_4.A_4=0 \: \: \: \: \: \: \: \: \: \: \: \: \: \: \: \: \: \: \: \: \: \: \: \: \: \: \: \: \: \: \nonumber \\
B_4.B_4.B_4.A_4+B_4.B_4.A_4.B_4+B_4.A_4.B_4.B_4+A_4.B_4.B_4.B_4=0 \: \: \: \: \: \: \: \: \: \: \: \: \: \: \: \: \: \: \: \: \: \: \: \: \: \: \: \: \: \: \nonumber \\
B_4.B_4.B_4.Q_4+B_4.B_4.Q_4.B_4+B_4.Q_4.B_4.B_4+Q_4.B_4.B_4.B_4=0 \: \: \: \: \: \: \: \: \: \: \: \: \: \: \: \: \: \: \: \: \: \: \: \: \: \: \: \: \: \: \nonumber \\
Q_4.Q_4.Q_4.A_4+Q_4.Q_4.A_4.Q_4+Q_4.A_4.Q_4.Q_4+A_4.Q_4.Q_4.Q_4=0 \: \: \: \: \: \: \: \: \: \: \: \: \: \: \: \: \: \: \: \:  \: \: \: \: \: \: \: \: \: \: \nonumber \\
Q_4.Q_4.Q_4.B_4+Q_4.Q_4.B_4.Q_4+Q_4.B_4.Q_4.Q_4+B_4.Q_4.Q_4.Q_4=0 \: \: \: \: \: \: \: \: \: \: \: \: \: \: \: \: \: \: \: \: \: \: \: \: \: \: \: \: \: \: \nonumber \\
A_4.A_4.B_4.B_4+B_4.B_4.A_4.A_4+A_4.B_4.A_4.B_4+B_4.A_4.B_4.A_4+A_4.B_4.B_4.A_4+B_4.A_4.A_4.B_4=0 \nonumber \\
A_4.A_4.Q_4.Q_4+Q_4.Q_4.A_4.A_4+A_4.Q_4.A_4.Q_4+Q_4.A_4.Q_4.A_4+A_4.Q_4.Q_4.A_4+Q_4.A_4.A_4.Q_4=0 \nonumber \\
B_4.B_4.Q_4.Q_4+Q_4.Q_4.B_4.B_4+B_4.Q_4.B_4.Q_4+Q_4.B_4.Q_4.B_4+B_4.Q_4.Q_4.B_4+Q_4.B_4.B_4.Q_4=0\nonumber \\
A_4.B_4.Q_4.Q_4+A_4.Q_4.B_4.Q_4+Q_4.A_4.B_4.Q_4+A_4.Q_4.Q_4.B_4+Q_4.A_4.Q_4.B_4+Q_4.Q_4.A_4.B_4 \: \: \: \: \: \: \: \: \: \: \nonumber \\
+B_4.A_4.Q_4.Q_4+B_4.Q_4.A_4.Q_4+Q_4.B_4.A_4.Q_4+B_4.Q_4.Q_4.A_4+Q_4.B_4.Q_4.A_4+Q_4.Q_4.B_4.A_4 = 0 \nonumber \\
B_4.Q_4.A_4.A_4+B_4.A_4.Q_4.A_4+A_4.B_4.Q_4.A_4+B_4.A_4.A_4.Q_4+A_4.B_4.A_4.Q_4+A_4.A_4.B_4.Q_4 \: \: \: \: \: \: \: \: \: \:  \nonumber \\
+ Q_4.B_4.A_4.A_4+Q_4.A_4.B_4.A_4+A_4.Q_4.B_4.A_4+Q_4.A_4.A_4.B_4+A_4.Q_4.A_4.B_4+A_4.A_4.Q_4.B_4 = 0 \nonumber \\
A_4.Q_4.B_4.B_4+A_4.B_4.Q_4.B_4+B_4.A_4.Q_4.B_4+A_4.B_4.B_4.Q_4+B_4.A_4.B_4.Q_4+B_4.B_4.A_4.Q_4 \: \: \: \: \: \: \: \: \: \: \nonumber \\
+Q_4.A_4.B_4.B_4+Q_4.B_4.A_4.B_4+B_4.Q_4.A_4.B_4+Q_4.B_4.B_4.A_4 \nonumber \\
+B_4.Q_4.B_4.A_4+B_4.B_4.Q_4.A_4= 0 \: \: \: \: \: \: \: \: \: \: 
\end{eqnarray}
In particular, we choose $Q_4$ to be the clock matrix, the most obvious generalization of $\sigma_3$, i.e., with $z_4=e^{i \frac{2 \pi}{4}}$
\begin{eqnarray}
Q_4 = \left(\begin{array}{cccc} 1 & 0 & 0 & 0  \\
						0 & z_4 & 0 & 0 \: \\
						0 & 0 & z_4^2 & 0 \\
						0 &  & 0 & z_4^3 \\  \end{array}\right) = \left(\begin{array}{cccc} 1 & 0 & 0 & 0  \\
						0 & i & 0 & 0 \: \\
						0 & 0 & -1 & 0 \\
						0 & 0  & 0 & -i \\  \end{array}\right)
\end{eqnarray}

There are several triplets of matrices that satisfy these conditions, as outlined in Dattoli {\it et al}\cite{Dattoli1}. A choice for $A$ and $B$ are
\begin{eqnarray}
A=\left(\begin{array}{cccc} 0 & -1 & 0 & 0  \\
						0 & 0 & 1 & 0 \: \\
						0 & 0 & 0 & -1 \\
						1 & 0 & 0 &0 \\  \end{array}\right) \: \: \: \: \: B^{\dagger}=e^{-  \frac{i \pi}{4}}\left(\begin{array}{cccc} 0 & -1 & 0 & 0  \\
						0 & 0 & -i & 0 \: \\
						0 & 0 & 0 & 1 \\
						i & 0 & 0 &0 \\  \end{array}\right)  \: \: \: \: \: 
\end{eqnarray}
In addition, in the above,
\begin{eqnarray}
T_{4+} =\left(\begin{array}{cccc} 0 & 0 & 1 & 0  \\
						0 & 0 & 0 & 1 \: \\
						0 & 0 & 0 & 0 \\
						0 & 0 & 0 &0 \\  \end{array}\right) \: \: \: \: \: T_{4-}=\left(\begin{array}{cccc} 0 & 0 & 0 & 0  \\
						0 & 0 & 0 & 0 \: \\
						1 & 0 & 0 & 0 \\
						0 & 1 & 0 &0 \\  \end{array}\right)  \: \: \: \: \: \: \: \: \: \: \: \: \: \: \: \: \: \: \: \: \: \: \: \: \: \: \: \: \: \: \: \: \: \: \: \: \: \: \: \: \: \: \: \: \: \: \: \: 
\end{eqnarray}
while
\begin{eqnarray}
a =  \sqrt{\sqrt{E^2-p_-p_+} -\sqrt{p_-p_+}} \: \: \:  \: \: \: \: \:  \: \: \: \: \:  \: \: 
W_4 =  \left(\begin{array}{cccc}a & 0 & 0 & 0  \\
						0 & a & 0 & 0 \: \\
						0 & 0 & -a & 0 \\
						0 &  & 0 & - a \\  \end{array}\right)  \: \: \: \: \: \: \: \: \: \: \:  \: \: \: \: \: \: \: \: \: \: \:  \: \: \: \: \: \: \: \: \: \: \:  \: \: \: \: \: \: \: \: \: \: \: \nonumber \\
b =  \sqrt{\sqrt{E^2-p_1^2-p_2^2} + \sqrt{E^2-p_1^2}} \: \: \:  \: \: \: \: \:  \: \: \: \: \:  \: \: Y_4 = \left(\begin{array}{cccc} b & 0 & 0 & 0  \\
						0 & b & 0 & 0 \: \\
						0 & 0 & - b & 0 \\
						0 &  & 0 & -b \\  \end{array}\right)   \: \: \: \: \: \: \: \: \: \: \:  \: \: \: \: \: \: \: \: \: \: \:  \: \: \: \: \: \: \: \: \: \: \:  \: \: \: \: \: \: \: \: \: \: \:\nonumber \\
c =  \sqrt{\sqrt{E^2-p_1^2-p_2^2} + \sqrt{E^2-p_2^2}} \: \: \:  \: \: \: \: \:  \: \: \: \: \:  \: \:Z_4 = \left(\begin{array}{cccc}c & 0 & 0 & 0  \\
						0 & c & 0 & 0 \: \\
						0 & 0 & - c & 0 \\
						0 &  & 0 & - c \\  \end{array}\right)  \: \: \: \: \: \: \: \: \: \: \:  \: \: \: \: \: \: \: \: \: \: \:  \: \: \: \: \: \: \: \: \: \: \:  \: \: \: \: \: \: \: \: \: \: \:
\end{eqnarray}
Note the useful relations
\begin{eqnarray}
T_{4+}T_{4-}+T_{4-} T_{4+}={\cal I}_4  \: \: \: \: \:  \: \: \: \: \:  \: \: \: \: \:  T_{4+}^2=T_{4-}^2=0
\end{eqnarray}

We have written the equations in this fashion because in each separate basis, the appropriate transformation matrix (one rotation and two boosts) are simple diagonal matrices.

Additionally, in all the four cases above, the eigenvalues of the matrices on the right side of Equation (13) are $\pm \sqrt{m}$.
The four equations are related by similarity transformations that are not unitary. Changes in basis need not occur with unitary transformations, all we need is that the time evolution represented by the equations of motion is unitary.

In the representation of the problem as in the second line in Equation (13), the effect of rotation is $p_{\pm} \rightarrow p_{\pm} e^{\pm i \theta}$, while $E \rightarrow E, m\rightarrow m$ are unchanged. Then, applying the following transformation (${\cal G}_R$ below)  renders the form of the equation invariant.
\begin{eqnarray}
R_4 = e^{\frac{i}{4}\theta  \: {\cal G}_R}  \: \: \: \: \: \: \: \: \: \: \: \: {\cal G}_R = \left(\begin{array}{cccc} -\frac{3}{2} & 0 & 0 & 0 \\
						0 & -\frac{1}{2} & 0 & 0 \\
						0 & 0 & \frac{1}{2} & 0   \\
						0 & 0 & 0 & \frac{3}{2} \: \\  \end{array}\right) \: \: \: \: \: \: \: \:
\rightarrow \: \: \: \: \: R_4 =	\left(\begin{array}{cccc} e^{-\frac{3i}{8} \theta} & 0 & 0 & 0  \\
						0 & e^{-\frac{i}{8} \theta} & 0 & 0 \\
						0 & 0 &  e^{\frac{i}{8} \theta}    & 0 \\
						0 & 0 & 0 &e^{\frac{3i}{8} \theta}    \\  \end{array}\right)
\end{eqnarray}
This equation thus represents spin-$\frac{3}{8}$ and spin-$\frac{1}{8}$ particles, since that captures the number of multiples of $2 \pi$ to render the components to 1.

Meanwhile, in the representation of the problem as in the third line of Equation (13), the effect of a boost (denoted by the rapidity $t$) along the $1$-direction is $(E-p_1) \rightarrow (E-p_1) e^{-t} \: , \: (E+p_1) \rightarrow (E+p_1) e^t$, while $p_2 \rightarrow p_2 \: , \: m \rightarrow m$ are unchanged. Then, in this representation, the following transformation (${\cal  G}_B^{(1)}$) renders the form of the equation invariant.
\begin{eqnarray}
R_4^B = e^{\frac{1}{4}t  \:{\cal G}_B^{(1)}}  \: \: \: \: \: \: \: \: \: \: \: \: {\cal G}_B^{(1)} = \left(\begin{array}{cccc} -\frac{3}{2} & 0 & 0 & 0 \\
						0 & -\frac{1}{2} & 0 & 0 \\
						0 & 0 & \frac{1}{2} & 0   \\
						0 & 0 & 0 & \frac{3}{2} \: \\  \end{array}\right)\: \: \: \: \: \: \: \: 
\rightarrow \: \: \:  R_4^B =	\left(\begin{array}{cccc} e^{-\frac{3}{8} t} & 0 & 0 & 0  \\
						0 & e^{-\frac{1}{8} t} & 0 & 0 \\
						0 & 0 &  e^{\frac{1}{8} t}    & 0 \\
						0 & 0 & 0 &e^{\frac{3}{8} t}    \\  \end{array}\right)	
\end{eqnarray}
Again, Equation (16) thus represents spin-$\frac{3}{8}$ and spin-$\frac{1}{8}$ particles.

Using the recipe  $E \rightarrow i \hbar \frac{\partial }{\partial t}, p_i \rightarrow -i \hbar \frac{\partial }{\partial x^{(i)}}$, we need to interpret fractional powers of these derivatives to represent fractional derivatives, which would be completely consistent with the application of these derivatives to plane-wave states. Accordingly, we write the equation for these particles using fractional derivatives \cite{frac} and where $\ket{\psi}$ is a $4$-component wave-function.
\begin{eqnarray}
m^{\frac{1}{2}} {\cal I}_4 \ket{\psi}= Q \: ( i \hbar \frac{\partial }{\partial t})^{\frac{1}{2}} + \sqrt{i} \left[ A (-i \hbar \frac{\partial }{\partial x^{(1)}})^{\frac{1}{2}} + B (-i \hbar \frac{\partial }{\partial x^{(2)}})^{\frac{1}{2}} \right] \ket{\psi}
\end{eqnarray}

In Section VI, we demonstrate that this equation describes unitary time-evolution.

\section{Generalization to higher even $N$}

Generalizing this to even $N$, we find the corresponding equations \cite{Satish2} yield spin-$\frac{2(N-1)}{N^2}$,$\frac{2(N-3)}{N^2}$, ...,$\frac{2}{N^2}$ particles.

\section{Generalizing to higher dimensional matrices and odd-fractional powers of variable, N=3}

The generalization to odd-N is a little different, since it is clear upon some reflection that the even-$N$ procedure would not give appropriate terms.
Starting with the energy-momentum relation, we write (${\cal I}_3$ is the $3 \times 3$ identity matrix) in the following three different ways, labeled as previously.
\begin{eqnarray}
Basis \: 2:  m^{\frac{1}{3}} {\cal I}_3 =W_3 + i^{\frac{1}{3}} \left[  T_{3+} p_-^{\frac{1}{3}} +  T_{3-} p_+^{\frac{1}{3}} + T_0 (p_- p_+)^{\frac{1}{6}} \right] \: \: \: \: \: \: \: \: \: \: \: \: \: \:\: \: \: \: \:\: \: \: \: \:\: \: \:    \\
Basis \: 3: m^{\frac{1}{3}} {\cal I}_3 = i^{\frac{1}{3}} Y_3 +  \left[  T_{3+} (E-p_1)^{\frac{1}{3}} +  T_{3-} (E+p_1)^{\frac{1}{3}}+ T_0 ((E-p_1) (E-p_1))^{\frac{1}{6}} \right] \nonumber \\
Basis \: 4:  m^{\frac{1}{3}} {\cal I}_3 = i^{\frac{1}{3}} Z_3 +  \left[  T_{3+} (E-p_2)^{\frac{1}{3}} +  T_{3-} (E+p_2)^{\frac{1}{3}} +T_0 ((E-p_2) (E-p_2))^{\frac{1}{6}} \right] \nonumber 
\end{eqnarray}
where we need to raise the two sides to the $6^{th}$ power to reproduce the usual energy-momentum relation. In the above,
\begin{eqnarray}
W_3 = \left(\begin{array}{ccc} \sqrt{m^{\frac{2}{3}} - i^{\frac{2}{3}} (p_- p_+)^{\frac{1}{3}} } & 0 & 0   \\
						0 & m^{\frac{1}{3}} - i^{\frac{1}{3}} (p_- p_+)^{\frac{1}{6} }& 0 \: \\
						0 &  0 & - \sqrt{m^{\frac{2}{3}} - i^{\frac{2}{3}} (p_- p_+)^{\frac{1}{3}} } \\  \end{array}\right) \: \: \: \: \: \: \: \: \nonumber \\
Y_3 = \left(\begin{array}{ccc} \sqrt{m^{\frac{2}{3}} - i^{\frac{2}{3}} (E^2-p_1^2)^{\frac{1}{3}} } & 0 & 0   \\
						0 & m^{\frac{1}{3}} - i^{\frac{1}{3}} (E^2-p_1^2)^{\frac{1}{6} }& 0 \: \\
						0 &  0 & - \sqrt{m^{\frac{2}{3}} - i^{\frac{2}{3}} (E^2-p_1^2)^{\frac{1}{3}} } \\  \end{array}\right) \: \: \: \: \: \: \: \: \nonumber \\
Z_3 = \left(\begin{array}{ccc} \sqrt{m^{\frac{2}{3}} - i^{\frac{2}{3}} (E^2-p_2^2)^{\frac{1}{3}} } & 0 & 0   \\
						0 & m^{\frac{1}{3}} - i^{\frac{1}{3}} (E^2-p_2^2)^{\frac{1}{6} }& 0 \: \\
						0 &  0 & - \sqrt{m^{\frac{2}{3}} - i^{\frac{2}{3}} (E^2-p_2^2)^{\frac{1}{3}} } \\  \end{array}\right) \: \: \: \: \: \: \: \: 
\end{eqnarray}
and
\begin{eqnarray}
T_{3+} =\left(\begin{array}{ccc} 0 & 0 & 1   \\
						0 & 0 & 0  \\
						0 & 0 & 0  \\  \end{array}\right) \: \: \: \: \: T_{3-}=\left(\begin{array}{ccc} 0 & 0 & 0   \\
						0 & 0 & 0 \\
						1 & 0 & 0  \\  \end{array}\right)  \: \: \: \: \: \: \: T_{0}=\left(\begin{array}{ccc} 0 & 0 & 0   \\
						0 & 1 & 0 \\
						0 & 0 & 0  \\  \end{array}\right)  \: \: \: \: \: \: \: \: \: \: \: \: \: \: \: \: \: \: \: \: \: \: \: \: \: \: \: \: \: \: \: \: \: \: \: \: \: \: \: \: \:
\end{eqnarray}

A little reflection shows that these represent spin-$\frac{1}{6}$ and spin $0$ particles. The appropriate rotation operator that leaves the first of the equations in Equation (24) invariant is
\begin{eqnarray}
R_3 = e^{\frac{i}{6}t  \:{\cal G}_R^{(1)}}  \: \: \: \: \: \: \: \: \: \: \: \: {\cal G}_R^{(1)} = \left(\begin{array}{ccc} -1 & 0 & 0 \\
						0 & 0 & 0 \\
						0 & 0 &  1 \: \\  \end{array}\right)\: \: \: \: \: \: \: \: 
\rightarrow \: \: \:  R_3 =	\left(\begin{array}{ccc} e^{-\frac{i}{6} t} & 0 & 0   \\
						0 & 1 & 0 \\
						0 & 0  &e^{\frac{i}{6} t}    \\  \end{array}\right)	
\end{eqnarray}
and a similar procedure applied to ``Basis 3'' allows us to deduce the boost matrix along axis-1, i.e.,
\begin{eqnarray}
R_3^B = e^{\frac{1}{6}t  \:{\cal G}_B^{(1)}}  \: \: \: \: \: \: \: \: \: \: \: \: {\cal G}_B^{(1)} = \left(\begin{array}{ccc} -1 & 0 & 0 \\
						0 & 0 & 0 \\
						0 & 0 &  1 \: \\  \end{array}\right)\: \: \: \: \: \: \: \: 
\rightarrow \: \: \:  R_3^B =	\left(\begin{array}{ccc} e^{-\frac{1}{6} t} & 0 & 0   \\
						0 & 1 & 0 \\
						0 & 0  &e^{\frac{1}{6} t}    \\  \end{array}\right)	
\end{eqnarray}

\section{Generalization to higher odd  $N$}

Generalizing this to larger odd-$N$, we find the corresponding equations \cite{Satish2} yield spin-$0, \frac{1}{(2 p-1) p}, \frac{2}{(2 p-1) p},...$ particles, where $N = 2 p -1$.

Interestingly, this approach does not appear to lead to an equation for spin-$\frac{1}{3}$ particles.

\section{The time evolution is unitary}

We study the time evolution using one of the formulations, for the specific case of $N=4$, using the usual rules of fractional calculus, which can be found in introductory references \cite{frac}. Essentially, we recognize,
\begin{eqnarray}
\frac{\partial^{\frac{1}{2}}}{\partial t^{\frac{1}{2}}} \: \frac{\partial^{\frac{1}{2}}}{\partial t^{\frac{1}{2}}} = \frac{\partial }{\partial t}
\end{eqnarray}
Using this, we can start with the first equation in Equation (14) and write, after a little re-arrangement
\begin{eqnarray}
E^{\frac{1}{2}} =  m^{\frac{1}{2}}  Q_4^{-1} - \sqrt{i} \left[ Q_4^{-1} B p_1^{\frac{1}{2}} + Q_4^{-1}A p_2^{\frac{1}{2}} \right] 
\end{eqnarray}
It can be easily checked the matrices $Q_4^{-1}, Q_4^{-1}A, Q_4^{-1}B$ all 4-anti commute, so that raising both sides to the $4^{th}$ power reproduces the relation $E^2=m^2+p_1^2+p_2^2$.

Now, making the replacement $E \rightarrow i \frac{\partial }{\partial t}$ and interpreting $E^{1/2} \rightarrow \frac{\partial^{\frac{1}{2}}}{\partial t^{\frac{1}{2}}}$, we get
\begin{eqnarray}
\sqrt{i} \frac{\partial^{\frac{1}{2}}}{\partial t^{\frac{1}{2}}} \psi = {\cal A} \psi
\end{eqnarray}
where $\cal A$ is the operator on the r.h.s. of Equation (32). Note that ${\cal A}^4=m^2+p_1^2+p_2^2$. Now, we can apply the fractional time derivative again to get
\begin{eqnarray}
i \frac{\partial }{\partial t} \psi = {\cal A}^2 \psi
\end{eqnarray}
and the operator ${\cal A}^2$ has strictly real eigenvalues ($\pm (m^2+p_1^2+p_2^2)$). Equation (34) thus describes unitary time-evolution.

\section{Conclusions}
We have constructed unitary time-evolution equations for particles with fractional spin in (2+1)-dimensions and shown explicitly that they are invariant to boosts and rotations. These equations involve fractional derivatives of $t$ and $x_1, x_2$. Extensions have been made for even-roots (4th, 6th etc.) and odd-roots (3rd, 5th etc.). We then show that the time-evolution of the equation is unitary.

\section{Data Availability Statement}
Data sharing is not applicable to this article as no new data were created or analyzed in this study.

\section{Acknowledgments}
The hospitality and fertile intellectual atmosphere of the NHETC at Rutgers is gratefully acknowledged. Extensive discussions with Scott Thomas were extremely fruitful.

\appendix

\end{document}